\documentclass{PoS}                              % Documentstyle for PoS submission

\usepackage[latin1]{inputenc}                    % Codepage latin1
\usepackage{graphicx}                            % Graphics package
\usepackage{latexsym}                            % Load additional symbols
\usepackage{amsfonts}                            % Use AMS fonts,
\usepackage{amssymb}                             % symbols,
\usepackage{amsmath}                             % and definitions
\usepackage{dcolumn}                             % Align table columns
\usepackage{bm}                                  % Bold math

\graphicspath{{.}{Graphics/}{Figures/}}          % Path for graphics
\newcommand{\imag}{\mbox{i}}                     % Roman imaginary unit
\newcommand{\mps}{m_{\pi}}                       % m_\pi (pseudoscalar mass)
\title{Nucleon electromagnetic form factors with Wilson fermions}

\ShortTitle{Nucleon form factors}

\author{QCDSF/UKQCD Collaboration: M.~G{\"o}ckeler$^a$,
  Ph.~H{\"a}gler$^b$, R.~Horsley$^c$, Y.~Nakamura$^d$, M.~Ohtani$^a$,
  D.~Pleiter$^d$, P.~E.~L.~Rakow$^e$, A.~Sch{\"a}fer$^a$,
  G.~Schierholz$^{df}$, \speaker{W.~Schroers}$\,^d$
  \footnote{Current address: Department of Physics, 
    National Taiwan University, Taipei 10617, Taiwan},
  H.~St{\"u}ben$^g$, and J.~M.~Zanotti$^c$ \\
  \llap{$^a$}Institut f\"ur Theoretische Physik, Universit\"at
  Regensburg, 93040 Regensburg, Germany\\
  \llap{$^b$} Institut f{\"u}r Theoretische Physik T39,
  Physik-Department der TU M{\"u}nchen, James-Franck-Strasse,
  85747 Garching, Germany\\
  \llap{$^c$}School of Physics, University of Edinburgh,
  Edinburgh EH9 3JZ, UK\\
  \llap{$^d$}John von Neumann-Institut f\"ur Computing NIC /
  DESY, 15738  Zeuthen, Germany\\
  \llap{$^e$}Theoretical Physics Division, Department of Mathematical
  Sciences, University of Liverpool, Liverpool L69 3BX, UK\\
  \llap{$^f$}Deutsches Elektronen-Synchrotron DESY,
  22603 Hamburg, Germany\\
  \llap{$^g$}Konrad-Zuse-Zentrum f\"ur Informationstechnik Berlin,
  14195 Berlin, Germany}

\abstract{The nucleon electromagnetic form factors continue to be of
  major interest for experimentalists and phenomenologists alike. They
  provide important insights into the structure of nuclear matter. For
  a range of interesting momenta they can be calculated on the
  lattice. The limiting factor continues to be the value of the pion
  mass. We present the latest results of the QCDSF collaboration using
  gauge configurations with two dynamical, non-perturbatively improved
  Wilson fermions at pion masses as low as 350 MeV. \\ \\ DESY 07-153
  \\ Edinburgh 2007/23 \\ \\}

\FullConference{The XXV International Symposium on Lattice Field Theory\\
		July 30 - August 4 2007\\
		Regensburg, Germany}

\begin{document}

% --------------------------------------------------------------------------
%
%  Introduction
%
% --------------------------------------------------------------------------

\section{Introduction\label{sec:introduction}}
Protons and neutrons constitute most of the material world around
us. This makes them the most important particles subject to the strong
interaction. Their electromagnetic form factors were among the first
quantities investigated in hadron structure and they are known for
several decades. For a recent review on the status of experiment and
phenomenology consult~\cite{Punjabi:2005wq}. Nonetheless, despite all
the scrutiny they are still subject to surprises. Recently, a series
of experiments has been performed at Jefferson
Lab~\cite{Jones:1999rz}. These have revealed surprising new features
of the form factors --- namely, that their ratio $F_2(Q^2)/F_1(Q^2)$
does not behave as was expected from previous experiments and as
predicted by perturbative QCD\@. Resolving this mystery requires
model-independent non-perturbative methods. Lattice QCD provides such
a method without model assumptions.

A theoretical explanation of the behavior of the form factor ratio has
been suggested in~\cite{Belitsky:2002kj}. In fact, lattice
calculations have recently been found to show similar
behavior~\cite{Negele:2003ma,Gockeler:2006uu,Schroers:Bar07}. This is
an example of how lattice calculations can give important feedback to
phenomenology and experiment already today.

This paper focusses on the calculation of the behavior of the
electromagnetic vector and axial form factors of the isovector
combination $p-n$, i.e., the difference between up- and down-quarks
unless explicitely stated otherwise. Due to isospin symmetry the
disconnected contributions cancel in this situation and there is no
additional systematic error other than the usual lattice
systematics. From the Lorentz-invariant expansions of the vector and
the axial currents, we define the form factors via
\begin{eqnarray}
  \label{eq:ff-def}
  \langle N(\vec{p}')\vert J^\mu\vert N(\vec{p})\rangle &=&
  \bar{u}(\vec{p}')\left( \gamma^\mu F_1(Q^2) + \frac{\imag
      q_\alpha}{2 m_N}\sigma^{\alpha\mu} F_2(Q^2)\right) u(\vec{p})\,,
  \nonumber \\
  \langle N(\vec{p}')\vert A^\mu\vert N(\vec{p})\rangle &=&
  \bar{u}(\vec{p}')\left( \gamma^\mu G_A(Q^2) + \frac{q^\mu}{2 m_N}
    G_P(Q^2) \right) \gamma_5 u(\vec{p})\,,
\end{eqnarray}
with $\vec{p}'$ ($\vec{p}$) being the initial (final) nucleon
momentum, $q=p'-p$, $Q^2=-q^2$ the virtual momentum transfer, and
$m_N$ being the nucleon mass.

We employ two flavors of dynamical Wilson-clover fermions with pion
masses as low as $\mps=350$~MeV~\cite{Gockeler:2006ns}. The lattice
spacing varies between $a=0.065\dots0.08$~fm where the scale has been
set by $r_0=0.45$~fm. This technology has the advantage that it
extends the successful calculations of the past decade without
conceptual problems like square roots of sea quarks, flavor and/or
taste breaking, residual masses or absence of unitarity away from the
continuum limit. The technology to extract form factors has been
covered in detail in~\cite{Gockeler:2003ay,Gockeler:2003jfa}.

% --------------------------------------------------------------------------
%
%  Vector form factors
%
% --------------------------------------------------------------------------

\section{Vector form factors\label{sec:vector-form-factors}}
To investigate the large-distance behavior of the vector form factors
we adopt the following parameterizations of $F(Q^2)$ as functions of
$Q^2$, see also~\cite{Ohtani:2007sb}:
\begin{eqnarray}
  \label{eq:f12-params}
  F_1(Q^2) &=& \frac{A}{1+c_{12}Q^2+c_{14}Q^4}\,, \nonumber \\
  F_2(Q^2) &=& \frac{1+\kappa}{1+c_{22}Q^2+c_{26}Q^6}\,.
\end{eqnarray}
These parameterizations have been inspired
by~\cite{Belitsky:2002kj,Belushkin:2006qa} by neglecting logarithmic
corrections. Such perturbative logarithms are not expected to play a
role for the energy range $Q^2<4$~GeV$^2$ we are investigating. The
normalization factor, $A$, is used to fix the renormalization
constant. Due to charge conservation, we must have $F_1(Q^2)\equiv 1$
which fixes the operator renormalization. The
parameterizations~Eqs.~(\ref{eq:f12-params}) have a couple of
important properties:
\begin{itemize}
\item The fulfill the superconvergence relation
  \[ \int_{-4\mps^2}^{-\infty} dQ^2\, \mbox{Im}\, F_{1,2}(Q^2) = 0\,,
  \quad \int_{-4\mps^2}^{-\infty} dQ^2\, Q^2\, \mbox{Im}\, F_2(Q^2) =
  0\,. \] 
\item They fulfill the expected asymptotic behavior
  \[ F_1 \propto 1/(Q^2)^2\,,\quad F_2 \propto 1/(Q^2)^3\,. \]
\item They should exhibit an effective resonance pole for negative
  $Q^2$. In particular, we will verify below that the location of the
  pole lies at the vector meson mass. This serves as a consistency
  check between our lattice calculation and phenomenology.
\end{itemize}
Alternatively, one can use dipole and tripole-type fit
formulae~\cite{Gockeler:2006uu,Schroers:Bar07}. We believe, however,
that the parameterizations in Eqs.~(\ref{eq:f12-params}) describe the
physics better. It is important to point out that the lattice data
does not cover a sufficiently large range of $Q^2$ values to allow us
to favor one form over another. Hence, we require additional
phenomenological input and perform consistency checks by probing the
location of the poles of $F_1(Q^2)$ in Eq.~(\ref{eq:f12-params}).

From the parameterizations in Eq.~(\ref{eq:f12-params}) we can obtain
the charge radii via a Taylor expansion
\begin{equation}
  \label{eq:f12-chargerad}
  F_i(Q^2) = F_i(0)\left(1-\frac{1}{6}\langle r_i^2\rangle Q^2 + {\cal
      O}(Q^4)\right)\,,\quad \langle r_i^2\rangle = 6\, c_{i2}\,.
\end{equation}
It is evident that the radius only depends on $c_{12}$/$c_{22}$, but
not on the higher coefficients of the powers $Q^4$/$Q^6$.

\begin{figure}
  \centering
  \includegraphics[scale=0.3,clip=true]{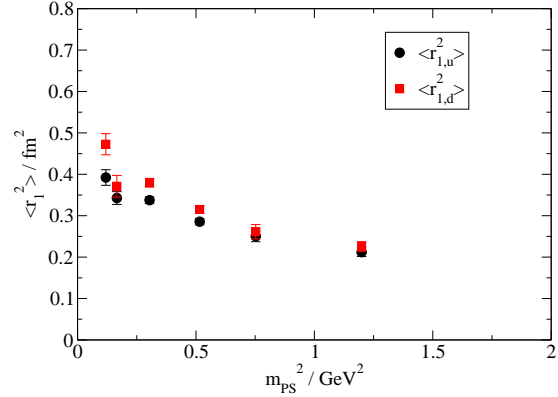}
  \caption{Difference in connected part of the charge radii for up-
    and down-quarks vs.~the pion mass.}
  \label{fig:flavor-dep}
\end{figure}

The first question we address is the flavor-dependence of the
connected contribution to the charge radius, $\langle
r_1^2\rangle$. Figure~\ref{fig:flavor-dep} shows the different charge
radii as a function of the pion mass at a fixed value of
$\beta=5.29$. For large pion masses the difference is below the
statistical uncertainty. As the pion mass decreases, however, the
difference starts to become significant. This is an excellent example
of why going towards sufficiently light quarks is crucial to study
phenomena related to hadron structure.

\begin{figure}
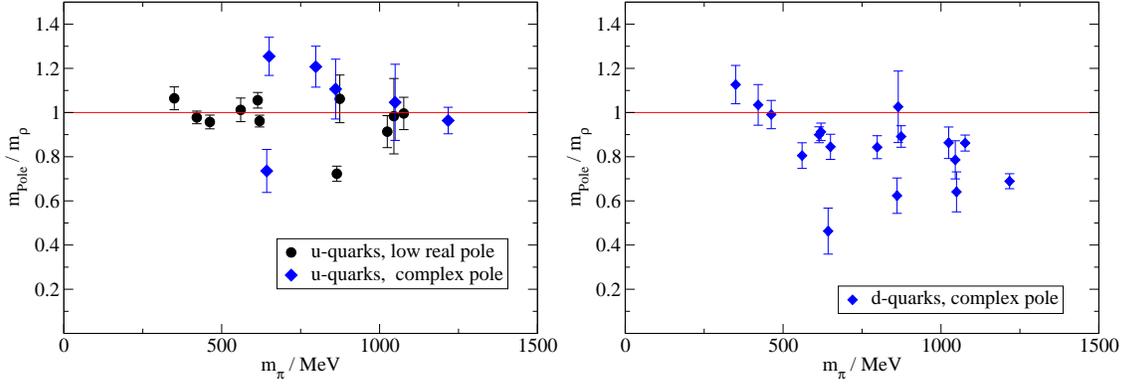

  \centering
  \includegraphics[scale=0.3,clip=true]{mpole_vs_mps-u-f1.eps}
  \includegraphics[scale=0.3,clip=true]{mpole_vs_mps-d-f1.eps}
  \caption{Location of the real part of the pole masses for u/d quarks
    (left/right panel). For u-quarks we find mostly two real poles and
    complex poles at a few working points. For d-quarks we find only
    complex poles.}
  \label{fig:poles}
\end{figure}

As it is evident from Eq.~(\ref{eq:f12-params}) the expression for
$F_1(Q^2)$ has in general two poles in the complex $Q^2$ plane. Both
poles can either be complex --- with a common real part --- or real
--- with a lower and an upper mass. From the vector meson dominance
model we expect the lower real pole and the real part of the complex
poles to be around the vector meson mass. Figure~\ref{fig:poles} shows
the location of the lower real/real part of the poles divided by the
vector meson mass at each working point. For u-quarks we find mostly
two real poles with a few complex poles at some working points. For
d-quarks we find complex poles exclusively. While the real poles are
indeed compatible with the vector meson mass, the real parts of the
complex poles scatter more strongly and tend to lie somewhat
lower. Understanding this phenomenon better would further improve our
picture of the nucleon.

\begin{figure}
  \centering
  \includegraphics[scale=0.3,clip=true]{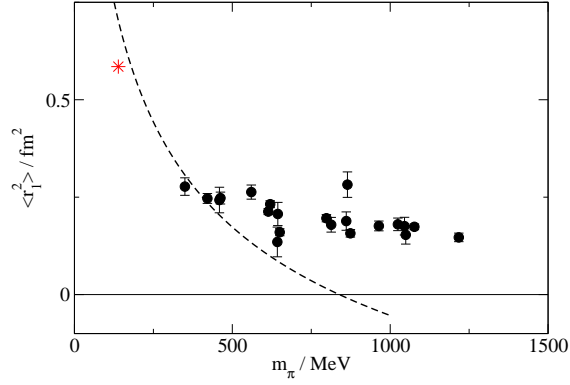}
  \caption{Isovector radius $\langle r_1^2\rangle$ as a function of
    the pion mass. The experimental value is denoted by a star, the
    dashed line is the expression obtained from the SSE.}
  \label{fig:r1v}
\end{figure}

The dependence of the charge radii on the pion mass has been studied
by different groups~\cite{Diehl:2006js,Gockeler:2003ay}. In the
following we focus on the small-scale expansion (SSE) given
in~\cite{Gockeler:2003ay}. Figure~\ref{fig:r1v} shows the lattice
results for $\langle r_1^2\rangle$ together with the experimental
point denoted by a star. The dashed curve is the SSE expression with
phenomenologically reasonable values for the parameters. The curve
grows to infinity as the pion mass goes to zero. Furthermore, it
vanishes at a finite value of the pion mass. The latter behavior is
unphysical, so we conclude that (not surprisingly) the SSE expression
at the order considered does not describe this quantity over the
entire range of pion masses available.

\begin{figure}
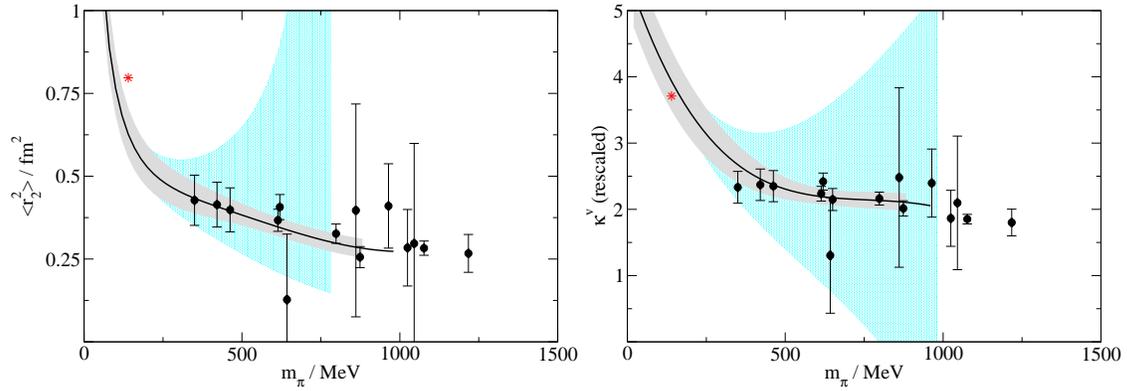

  \centering
  \includegraphics[scale=0.3,clip=true]{rv2Q.eps}
  \includegraphics[scale=0.3,clip=true]{kappav.eps}
  \caption{Isovector radius $\langle r_2^2\rangle$ (left panel) and
    anomalous magnetic moment $\kappa^v$ (right panel) from a combined
    fit of all lattice data. The rescaling has been explained in
    Ref.~\cite{Gockeler:2003ay}. The experimental values are denoted
    by stars. The error bands are statistical (shaded gray) and
    systematical (dotted cyan).}
  \label{fig:r2v-kappav}
\end{figure}

Figure~\ref{fig:r2v-kappav} shows the results of a combined fit of all
lattice data for $\langle r_2^2\rangle$ and $\kappa^v$. In these
expressions there are four parameters that we fit. The error bands
show the statistical errors and the systematic error due to higher
orders in the chiral expansion as determined by varying the individual
fit parameters and checking the stability of the
result~\cite{Meissner:2006pr}. For $\langle r_2^2\rangle$ the
extrapolation misses the experimental data point, while for $\kappa^v$
we find that the extrapolation is compatible with the experimental
data point. It is premature to conclude that the expansion fails for
$\langle r_2^2\rangle$ as it did for $\langle r_1^2\rangle$ since the
uncertainties of the input parameters may be
underestimated. Nonetheless, the range of applicability of the chiral
expansion is limited and further study is needed to accumulate
sufficiently accurate data points at sufficiently small pion masses.

% --------------------------------------------------------------------------
%
%  Axial form factors
%
% --------------------------------------------------------------------------

\section{Axial form factors\label{sec:axial-form-factors}}
Analogous to the form factors $F_1(Q^2)$ and $F_2(Q^2)$ the axial form
factor, $G_A(Q^2)$, and the induced pseudoscalar form factor,
$G_P(Q^2)$, can be calculated. For a review on experimental methods
and phenomenological parameterizations see~\cite{Bernard:2001rs}. The
axial form factor is usually fitted using a dipole ansatz
\begin{equation}
  \label{eq:ga-dipole}
  G_A(Q^2) = \frac{g_A}{(1+Q^2/M_A^2)^2}\,,
\end{equation}
with the axial coupling $g_A$ being one of the milestones of lattice
QCD calculations~\cite{Edwards:2005ym}, see
also~\cite{Schroers:2007pr}. The induced pseudoscalar coupling
$G_P(Q^2)$ is well understood and can be fitted using a pion-pole
ansatz for sufficiently small $\mps$ and $Q^2$,
see~\cite{Bernard:2001rs}:
\begin{equation}
  \label{eq:gp-pionpole}
  G_P(Q^2) = \frac{4 m_N g_{\pi N} f_\pi}{\mps^2+Q^2} - \frac{2}{3}
  g_A m_N^2\langle r_A^2\rangle + {\cal O}(Q^2,\mps^2)\,.
\end{equation}
We want to test this ansatz by verifying that the position of the pole
is exactly at the position of the pion mass at this working point.

\begin{figure}
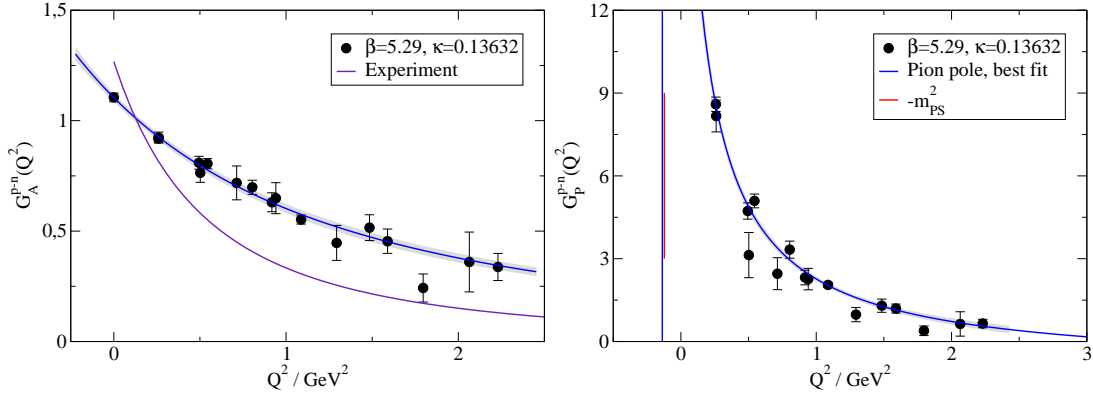

  \centering
  \includegraphics[scale=0.3,clip=true]{b5p29kp13632-GA.eps}
  \includegraphics[scale=0.3,clip=true]{b5p29kp13632-GP.eps}
  \caption{Isovector axial form factors for a sample working point at
    our lowest pion mass of $\mps=350$~MeV. The axial form factor
    (left panel) is fitted using a dipole form, the experimental best
    dipole fit is included. The pseudoscalar form factor (right panel)
    is fitted using a pion pole expression with the evaluated
    pseudoscalar mass being shown for comparison.}
  \label{fig:axialff}
\end{figure}

Our results for $G_A(Q^2)$ and $G_P(Q^2)$ are shown in
Fig.~\ref{fig:axialff} for a sample working point at our lowest pion
mass of $\mps=350$~MeV. The axial coupling fitted with a dipole is
compared to the corresponding experimental result. It is evident that
the lattice curve is flatter, implying that the heavy quarks on our
lattice build a smaller nucleon than Nature --- a phenomenon that has
been observed previously for the vector form factors,
cf.~Figs.~\ref{fig:r1v} and~\ref{fig:r2v-kappav}.
The induced pseudoscalar form factor indeed exhibits a pole at the
location of the measured pion mass --- as indicated in the plot by a
vertical bar. It is evident that the pole obtained from the fitted
pion-pole parameterization is fully consistent with the value of the
pion mass. Hence, our fit seems to be remarkably consistent with the
expression~Eq.~(\ref{eq:gp-pionpole}).

% --------------------------------------------------------------------------
%
%  Summary
%
% --------------------------------------------------------------------------

\section{Summary\label{sec:summary}}
We have successfully measured the vector and the axial vector form
factors. For the vector form factors a variety of chiral expansions is
available. The radius $\langle r_1^2\rangle$ cannot be described well,
while the situation for $\langle r_2^2\rangle$ and $\kappa^v$ is more
favorable. The axial form factor $G_A(Q^2)$ turns out to be described
well by a dipole formula --- just like the experimental data --- but
the curve is flatter indicating that also in this situation a proper
chiral extrapolation is essential. We verified that the induced
pseudoscalar form factor, $G_P(Q^2)$, is described excellently by the
pion pole picture already at $\mps=350$~MeV.

We conclude by pointing out that dynamical clover fermions provide a
viable technology to treat QCD with light quarks. The currently
running simulations may extend down to pion masses as low as
$\mps=250$~MeV within the next year. We are optimistic to reach
$\mps=200$~MeV by the end of this decade.

\acknowledgments The numerical calculations have been performed on the
Hitachi SR8000 at LRZ (Munich), the BlueGene/L and the Cray T3E at
EPCC (Edinburgh)~\cite{UKQCD}, the BlueGene/Ls at NIC/JFZ (J{\"u}lich)
and KEK (by the Kanazawa group as part of the DIK research program)
and on the APEmille and apeNEXT at NIC/DESY (Zeuthen). This work was
supported in part by the DFG under contract FOR 465 (Forschergruppe
Gitter-Hadronen-Ph{\"a}nomenologie and Emmy-Noether program) and by
the EU Integrated Infrastructure Initiative Hadron Physics (I3HP)
under contract number RII3-CT-2004-506078. W.S.~thanks Wolfgang
Bietenholz for valuable discussions.

% --------------------------------------------------------------------------
%
%  Bibliography
%
% --------------------------------------------------------------------------

% --------------------------------------------------------------------------
% 
%  End of document
% 
% --------------------------------------------------------------------------

\end{document}